\documentclass[
reprint,
superscriptaddress,
preprintnumbers,
 amsmath,amssymb,
 aps,showpacs,showkeys
]{revtex4-1}

\usepackage{graphicx}
\usepackage{dcolumn}
\usepackage{bm}

\begin{document}
\title{Finding the Resistance Distance and Eigenvector Centrality from the Network's Eigenvalues}
\author{Carac{\'e} Guti{\'e}rrez}
\affiliation{Universidad de la Rep{\'u}blica, Instituto de F{\'i}sica de Facultad de Ciencias, Igu{\'a} 4225, Montevideo 11400, Uruguay.}
\author{Juan Gancio}
\affiliation{Universidad de la Rep{\'u}blica, Instituto de F{\'i}sica de Facultad de Ciencias, Igu{\'a} 4225, Montevideo 11400, Uruguay.}
\author{Cecilia Cabeza}
\affiliation{Universidad de la Rep{\'u}blica, Instituto de F{\'i}sica de Facultad de Ciencias, Igu{\'a} 4225, Montevideo 11400, Uruguay.}
\author{Nicol{\'a}s Rubido} \email{nrubido@fisica.edu.uy}\email{nicolas.rubidoobrer@abdn.ac.uk}
\affiliation{Universidad de la Rep{\'u}blica, Instituto de F{\'i}sica de Facultad de Ciencias, Igu{\'a} 4225, Montevideo 11400, Uruguay.}
\affiliation{University of Aberdeen, King's College, Institute for Complex Systems and Mathematical Biology, AB24 3UE Aberdeen, United Kingdom.}
\date{\today}
\begin{abstract}
There are different measures to classify a network's data set that, depending on the problem, have different success. For example, the resistance distance and eigenvector centrality measures have been successful in revealing ecological pathways and differentiating between biomedical images of patients with Alzheimer's disease, respectively. The resistance distance measures the effective distance between any two nodes of a network taking into account all possible shortest paths between them and the eigenvector centrality measures the relative importance of each node in the network. However, both measures require knowing the network's eigenvalues and eigenvectors -- eigenvectors being the more computationally demanding task. Here, we show that we can closely approximate these two measures using only the eigenvalue spectra, where we illustrate this by experimenting on elemental resistor circuits and paradigmatic network models -- random and small-world networks. Our results are supported by analytical derivations, showing that the eigenvector centrality can be perfectly matched in all cases whilst the resistance distance can be closely approximated. Our underlying approach is based on the work by Denton, Parke, Tao, and Zhang [arXiv:1908.03795 (2019)], which is unrestricted to these topological measures and can be applied to most problems requiring the calculation of eigenvectors.
\end{abstract}
\maketitle

\section{Introduction}\label{sec_Into}
From human brain studies to social network analyses, we have seen enormous improvements in terms of data availability and accuracy. High-resolution brain fMRI scans are allowing us to study the brain's functional connectivity with increasing detail \cite{Sporns2009,Buldu2014,Fornito2016book}, which in turn, helps us to understand cognitive processes or the effects that diseases cause to the brain's connectivity. Social network platforms, such as Twitter and Facebook, produce unprecedented data streams, which, for example, allows us to study how information is shared among different communities \cite{Wasserman1994,DiazGuilera2003}. However, with an increasing ability to obtain larger and more reliable data sets, we also need to improve our data mining abilities in order to choose which content is the most important. In this mining process, network analysis has proven useful \cite{Zanin2016}.

Network science provides us with different measures to characterise, classify, and extract information from network data sets \cite{Newman2006book,Mendes2013book}. Namely, measurable quantities that highlight the most relevant relationships appearing within the data set. In general, the relevant measures are the invariant ones \cite{FanChung}, i.e., those that when nodes (elements in the data set) are relabelled the measure remains unchanged. In particular, the eigenvalue spectra and eigenvector set. These spectral characteristics have been shown to relate with other topological measures, such as the degree distributions \cite{Mendes2003,Mendes2004,Motter2007}, centrality measures \cite{Estrada2005,Pauls2012,Newman2014}, and modularity \cite{Wang2010,Newman2012,Peixoto2013}. 

An important measure to quantify distance between points (nodes) in a network is the resistance distance \cite{Klein1993,Zhang2007,Yang2007,Rubido2015book}, which is found from the network's eigenvalues and eigenvectors. This measure includes more information than the shortest-path joining two nodes, which is defined by the minimum number of edges (links) necessary to connect the nodes in a series path. The resistance distance also takes into account all other (non-repeating) paths connecting the two nodes to add them as parallel paths \cite{Klein1993,Rubido2015book,Asad2006,Asad2014,Rubido2017}. Hence, shortest-paths are relevant when there is a known corpuscular communication along the edges of the network, whilst the resistance distance is relevant when communication along the edges is spread like a wave pattern (as in extended systems). For example, the resistance distance has been used to detect communities (i.e., group nodes into modules) \cite{Girvan2002,Rubido2013,Zhang2019}, explain transport phenomena \cite{Stanley2005,Lai2009,Katifori2010,Rubido2014}, describe stochastic growth processes \cite{Korniss2004,Korniss2005,Korniss2007}, and reveal gene flows \cite{Beier2007} or ecological pathways \cite{Bowman2017,Thiele2018}.

Another relevant measure is the eigenvector centrality, which quantifies the relative importance of each node in a network and is based on the network's Perron-Frobenius eigenvector \cite{Chang2008}. This measure provides a score to each node according to the (positive) entries of the eigenvector associated to the largest eigenvalue of the adjacency matrix \cite{Newman2014,Recht2000,Bonacich2007,Sharkley2019}. This has been particularly useful in biomedical image analysis \cite{Turner2010} for a broad range of studies, for example, Alzheimer's disease \cite{Wink2014}, type-I diabetes \cite{Wink2017}, and ageing \cite{Zhang2017}. It has also been used for community detection \cite{Newman2006}, characterise protein pathways \cite{Negre2018}, and measure the elastic modulus of materials \cite{Welch2017}. Nevertheless, as with the resistance distance, the eigenvector centrality depends on finding eigenvectors, which for large networks can be computationally demanding.

Here we show that, by solely knowing the eigenvalue spectra of the network, we can approximate its resistance distance values and derive exactly its eigenvector centrality. Our method is based on the work by Denton, Parke, Tao, and Zhang \cite{Denton2019} (who recently recovered an algebraic relationship to find the magnitudes of eigenvector's components from the eigenvalue spectra). Our results include analytical derivations on the expressions for the resistance distance and eigenvector centrality measures as well as experimental and numerical examples. In particular, we analyse small-sized resistor circuit networks ($N\sim\mathcal{O}(10^1)$) and numerically generated random \cite{Erdos1960} and small-world networks \cite{Strogatz1998} ($N\sim\mathcal{O}(10^2)-\mathcal{O}(10^3)$). We note that, aside cases where the network has a degenerate eigenvalue spectra, our conclusions are general and unrestricted to these topological measures and examples.

\section{Methods}\label{sec_Methods}
Denton el al.~\cite{Denton2019} have recently shown that the components of eigenvectors can be recovered from the eigenvalue spectra, which they name as \emph{eigenvector-eigenvalue identity}, and is valid for any Hermitian matrix with non-degenerate eigenvalues. 
Specifically, the identity holds
\begin{equation}
    \left|[\vec{\psi}_n]_i\right|^2 = \frac{ \prod_{k = 1}^{N-1} \left[ \lambda_n(\mathbf{A}) - \lambda_k(\mathbf{M}_i)\right] }{ \prod_{k = 1;\,k \neq n}^{N} \left[ \lambda_n(\mathbf{A}) - \lambda_k(\mathbf{A})\right] },
    \label{eq_Tao}
\end{equation}
where $[\vec{\psi}_n]_i$ is the $i$-th component ($i = 1,\ldots,N$) of the $n$-th eigenvector of matrix $\mathbf{A}$ associated to the eigenvalue $\lambda_n(\mathbf{A})$ (with $n = 1,\ldots,N$ modes), such that $\mathbf{A}\vec{\psi}_n = \lambda_n(\mathbf{A})\vec{\psi}_n$, and $\lambda_k(\mathbf{M}_i)$ is the $k$-th eigenvalue ($k = 1,\ldots,N-1$) of matrix $\mathbf{M}_i$, which is obtained from $\mathbf{A}$ by removing the $i$-th row and column. Also, without loss of generality, it can be assumed that the eigenvalue spectra in Eq.~\eqref{eq_Tao} is ordered non-decreasingly; that is, $\lambda_1(\mathbf{A}) \leq \lambda_2(\mathbf{A}) \leq \cdots \leq \lambda_N(\mathbf{A})$. Consequently, Eq.~\eqref{eq_Tao} allows to find the magnitudes of the eigenvector components from the eigenvalue spectra.

In this work, matrix $\mathbf{A}$ is the network's adjacency matrix. We restrict ourselves to undirected and unweighted networks, which correspond to having a binary, symmetric, adjacency matrix, $\mathbf{A}$. Namely, $A(i,j) = A(j,i)$ for all entries ($\mathbf{A} = \mathbf{A}^T$), where $A(i,j) = 1$ if there is an edge connecting nodes $i$ and $j$ in the network and $A(i,j) = 0$ otherwise. In particular, we use the Erd{\"o}s-R{\'e}nyi model \cite{Erdos1960} to generate random networks and the Watts-Strogatz model \cite{Strogatz1998} to generate small-world networks \cite{Numeric}. In order to find the network main characteristics, such as the average shortest-path and clustering coefficient, we use the Brain Connectivity Toolbox \cite{BCT}. 

We find the resistance distance between nodes $i$ and $j$, $\rho(i,j)$, by using the eigenvalues and eigenvectors of the network's Laplacian matrix, $\mathbf{L}$, as \cite{Klein1993,Zhang2007,Yang2007,Rubido2015book}
\begin{equation}
    \rho(i,j) = \sum_{n = 2}^N \frac{1}{\lambda_n(\mathbf{L})}\left|[\vec{\phi}_n]_i - [\vec{\phi}_n]_j\right|^2,
    \label{eq_ResistDist}
\end{equation}
where $\mathbf{L} = \mathbf{D} - \mathbf{A}$, $\mathbf{D}$ being the diagonal matrix containing all node degrees (i.e., the number of neighbours of each node) and $\mathbf{A}$ being the network's adjacency matrix. $\mathbf{L}\vec{\phi}_n = \lambda_n(\mathbf{L})\vec{\phi}_n$ for $n = 1,\ldots,N$. It is worth noting that when $\mathbf{A} = \mathbf{A}^T$ the Laplacian matrix is positive semi-defined \cite{FanChung}, implying that the eigenvalues, $\lambda_n(\mathbf{L})$, are always such that $0 = \lambda_1(\mathbf{L}) \leq \lambda_2(\mathbf{L}) \leq \cdots \leq \lambda_N(\mathbf{L})$.

\section{Results}\label{sec_Results}
Here, we derive an approximate value for the resistance distance [Eq.~\eqref{eq_ResistDist}] by constructing from Eq.~\eqref{eq_Tao} upper and lower bounds. In practical settings, where eigenvector calculations can be challenging, we show by different numerical and real-world networks that our approximation is well-defined. Then, we use Eq.~\eqref{eq_Tao} to exactly derive the eigenvector centrality measure, which implies that any difference in the numerical calculations appearing are due to numerical errors (such as round-off and truncation errors) and can be neglected.

The \emph{resistance distance}, $\rho$, requires knowing the eigenvalue spectra and all eigenvector components' magnitudes and signs. Since the eigenvalue-eigenvector identity of Eq.~\eqref{eq_Tao} only allows for the calculation of magnitudes, we are unable to know exactly how the differences in Eq.~\eqref{eq_ResistDist} are contributing to the overall $\rho(i,j)$ value. However, we can use the eigenvector magnitudes from Eq.~\eqref{eq_Tao} to find upper and lower bounds. Specifically, we use Eq.~\eqref{eq_Tao} to write the $i$-th component of $\vec{\psi}_n$ as
\begin{equation}
    \left|[\vec{\phi}_n]_i\right|^2 = \frac{ \prod_{k = 1}^{N-1} \left[ \lambda_n(\mathbf{L}) - \lambda_k(\mathbf{M}_i)\right] }{ \prod_{k = 1;\,k \neq n}^{N} \left[ \lambda_n(\mathbf{L}) - \lambda_k(\mathbf{L})\right] },
    \label{eq_LapEigenVecMag}
\end{equation}
where matrix $\mathbf{M}_i$ is now obtained from the network's Laplacian matrix, $\mathbf{L}$, by removing its $i$-th row and column. On the other hand, the difference between the $i$-th and $j$-th coordinates for the $n$-th mode in Eq.~\eqref{eq_ResistDist} is
\begin{equation}
    \left|[\vec{\phi}_n]_i - [\vec{\phi}_n]_j\right|^2 = \left|[\vec{\phi}_n]_i\right|^2 + \left|[\vec{\phi}_n]_j\right|^2 - 2 [\vec{\phi}_n]_i [\vec{\phi}_n]_j^*,
    \label{eq_LapEigenVecDiff}
\end{equation}
where the last cross-product term has an unknown sign -- it can be either positive or negative, depending on the signs of each eigenvector's component. Hence, we use Eq.~\eqref{eq_LapEigenVecMag} as follows. We do the summation in Eq.~\ref{eq_ResistDist} only using positive or negative cross-products terms, i.e., either we use $+2 | [\vec{\phi}_n]_i | | [\vec{\phi}_n]_j |$ or use $-2 | [\vec{\phi}_n]_i | | [\vec{\phi}_n]_j |$. When only adding [subtracting] these cross-product terms in Eq.~\eqref{eq_ResistDist} we are effectively generating an upper [a lower] bound for the resistance distance, which we note as $\rho_{up}(i,j)$ [$\rho_{down}(i,j)$]. Then, we use these bounds to approximate the exact $\rho(i,j)$ value by their average,
\begin{equation}
    \rho_{approx}(i,j) = \frac{1}{2}\left[\rho_{up}(i,j) + \rho_{down}(i,j)\right],\;\forall\,i,j.
    \label{eq_ApproxResistDist}
\end{equation}

\begin{figure}[htbp]
    \centering
    \includegraphics[width=0.49\columnwidth]{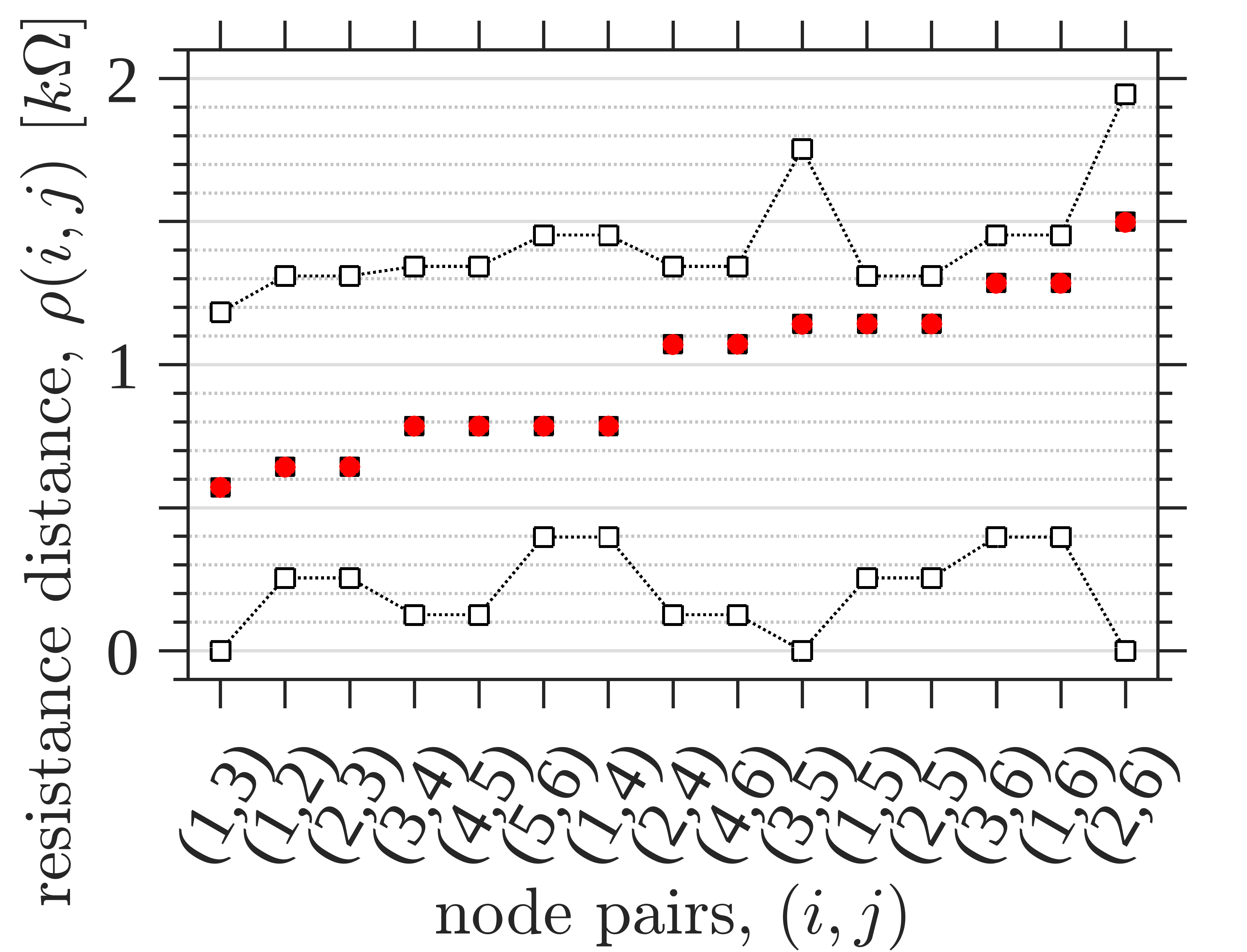}
    \includegraphics[width=0.49\columnwidth]{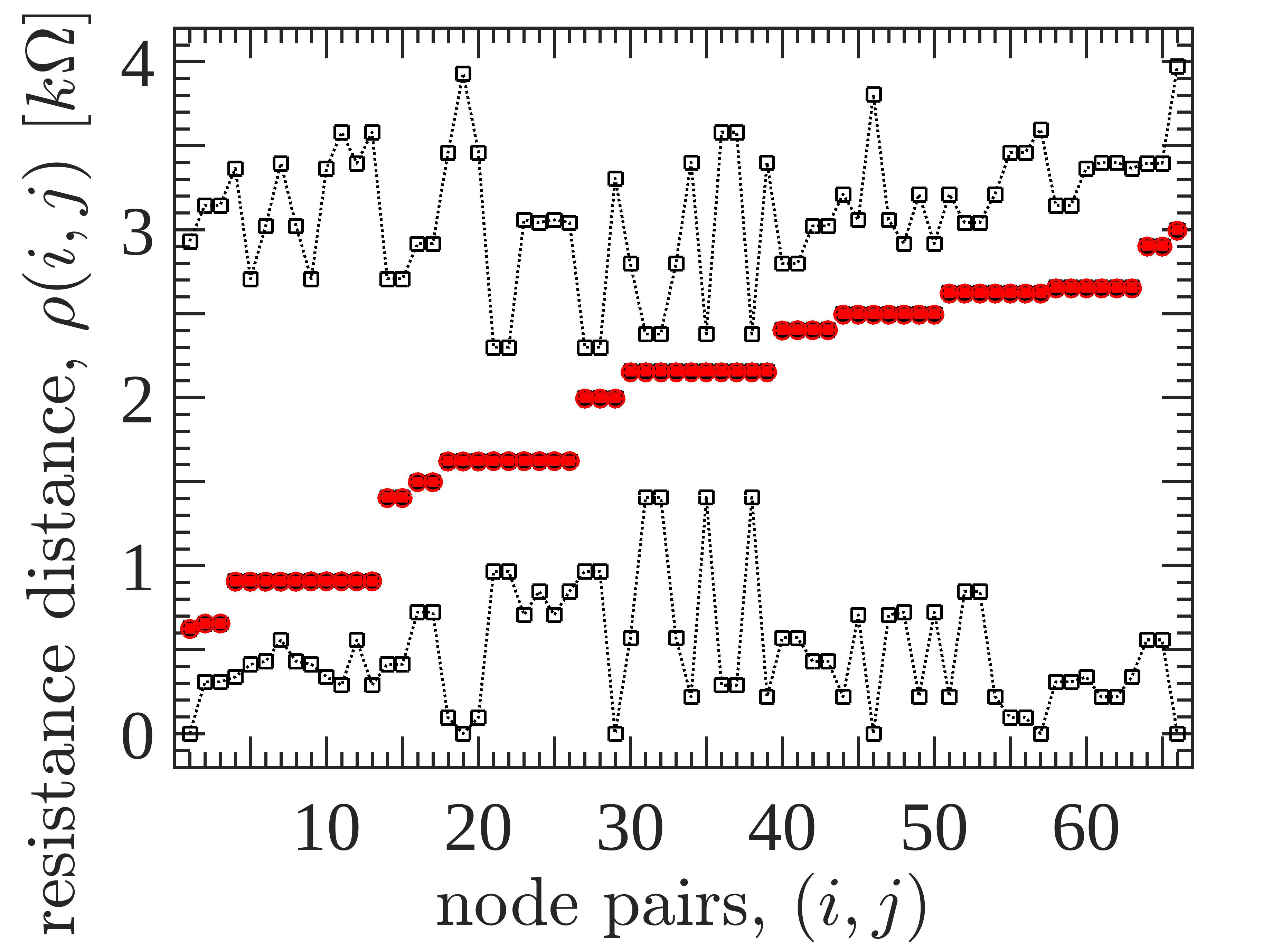}
    \includegraphics[width=0.49\columnwidth]{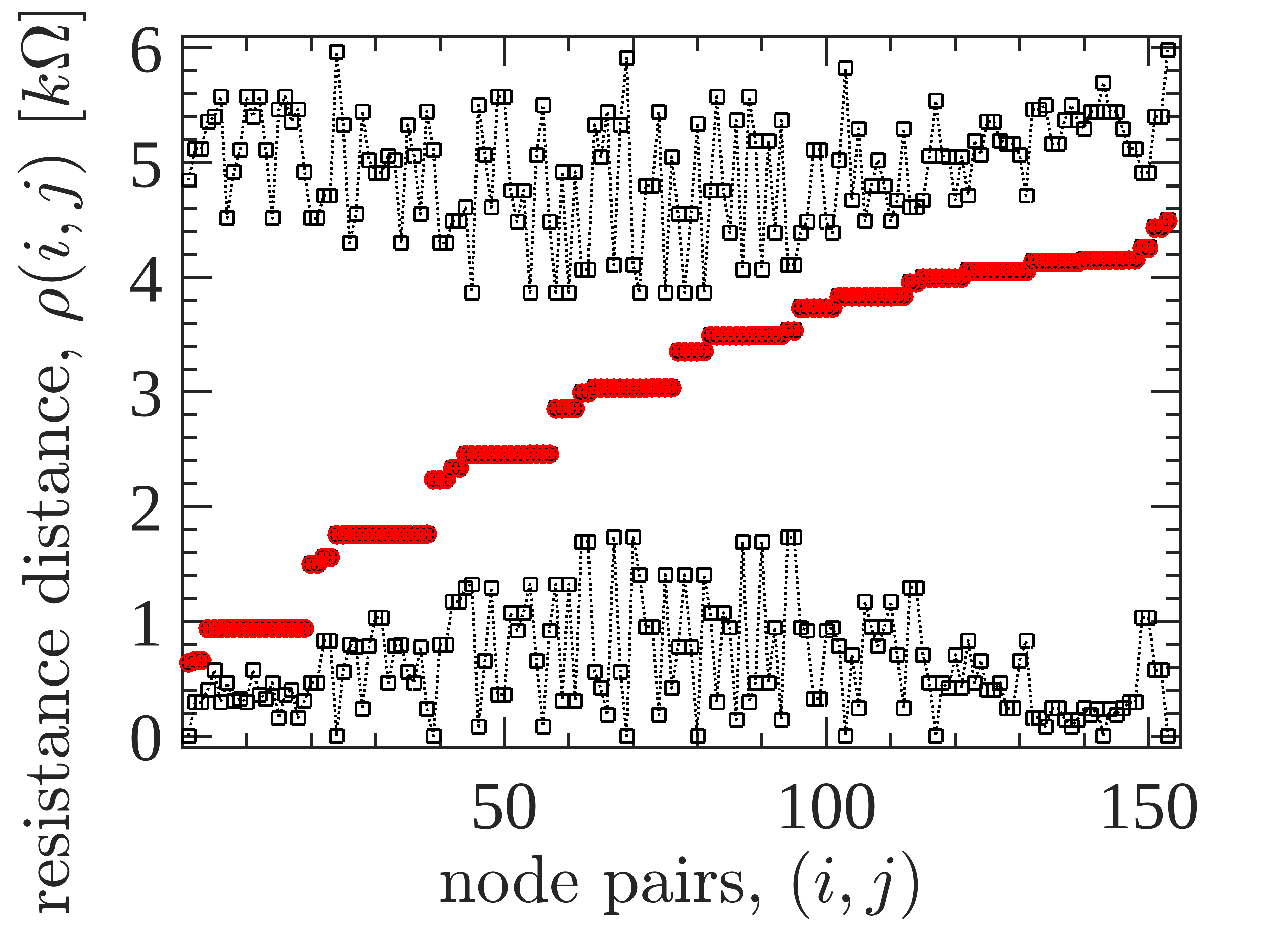}
    \includegraphics[width=0.49\columnwidth]{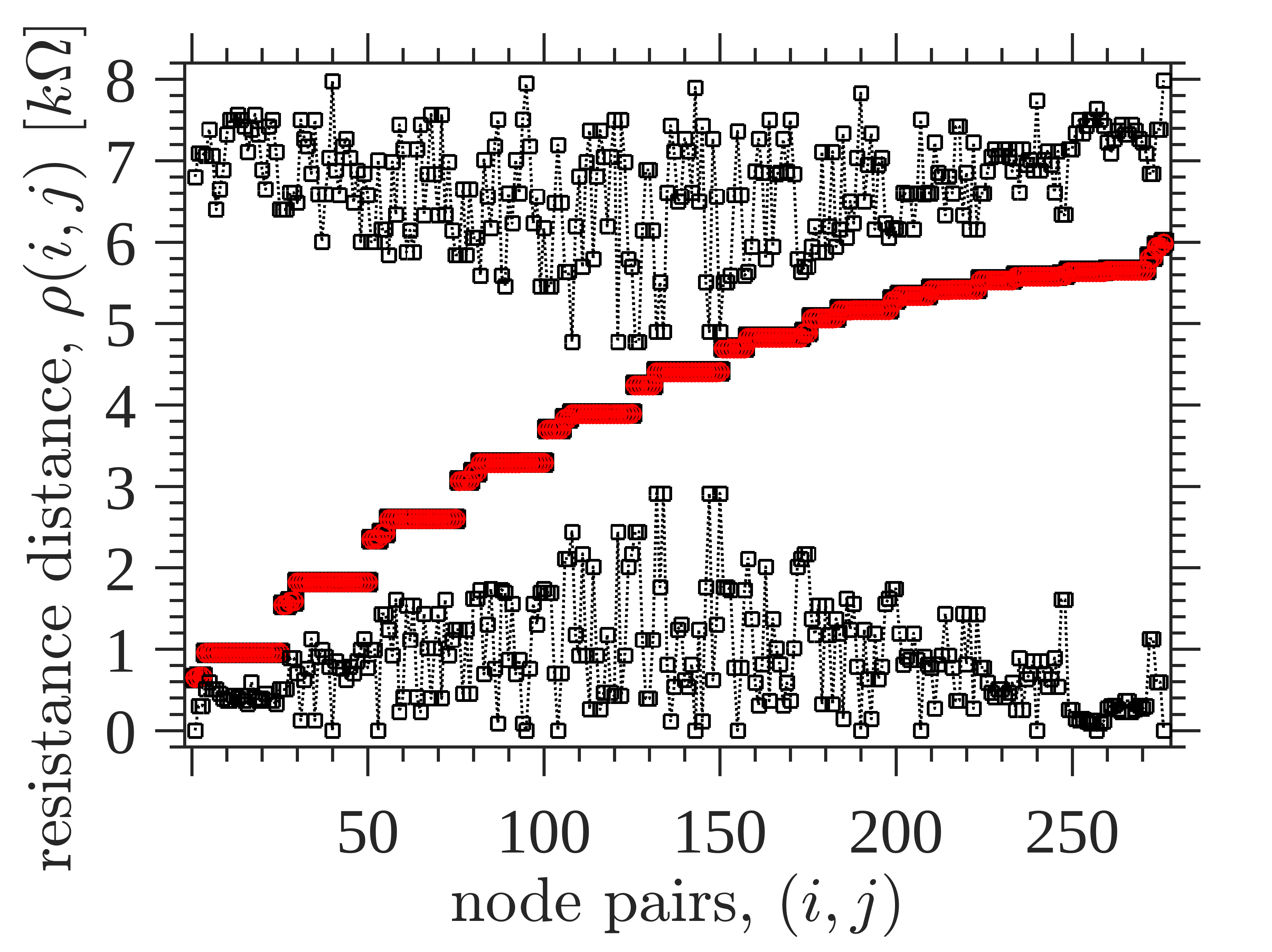}
    \caption{{\bf Resistance distance values and bounds for quasi-ring resistor networks}. These resistor networks are connected in a ring with an extra edge between $2$ non-neighbouring nodes, i.e., the edge $(1,3)$ is added. Filled (red online) circles represent the ohmmeter readings between each pair of nodes with the instruments resolution as error-bars (approximately $1\,\Omega$ [Ohm]), filled black squares show the corresponding theoretical resistance distance values, $\rho$ [Eq.~\eqref{eq_ResistDist}] (assuming a network with identical $1\,k\Omega$ resistors), and unfilled squares show our resistance distance bounds. From left to right and top to bottom, the number of nodes, $N$, is $6,\,12,\,18$, and $24$. All panels have been ordered according to ascending values of the measured $\rho$, where the top left panel shows specifically to which edges the values correspond, whilst the rest are indexed from $1$ up to $N(N-1)/2$ edges.}
    \label{fig_RingNetExp}
\end{figure}

The resistor networks in Fig.~\ref{fig_RingNetExp} are experimental resistors \cite{Asad2006,Asad2014} connected in quasi-ring structures. Namely, their adjacency matrix is such that, $A(i,j) = 1$ if $j = i\pm1$ and $0$ otherwise (when $i = 1$ and $j = N+1$, $A(N,N+1) = A(N,1) = 1$, and when $i = 1$ and $j = 0$, $A(1,0) = A(1,N)$), with an extra edge connecting nodes $1$ and $3$, i.e., $A(1,3) = 1 = A(3,1)$. We add this extra edge to the networks in order to lift the degenerate eigenvalues (i.e., repeated eigenvalues, spanning an eigenspace with more than $1$ eigenvector), which are always present in circulant networks \cite{FanChung,Yang2007}. From Fig.~\ref{fig_RingNetExp}, we can see that both bounds -- shown by unfilled symbols -- $\rho_{up}$ and $\rho_{down}$, exhibit a similar behaviour, which appears independently of $N$. Specifically, the intermediate values of $\rho$ -- measured from an ohmmeter or calculated from Eq.~\eqref{eq_ResistDist} (filled symbols) -- are closely bounded, whilst its extreme values (either for small $\rho(i,j)$ or for large $\rho(i,j)$) are loosely bounded by either $\rho_{up}$ or $\rho_{down}$, with one of the two bounds being closer to the measured $\rho$ in each opposite end. Meaning that, $\rho_{approx}$ follows the measured values (filled circles) closer than either of the bounds.

Specifically, the bounds in Fig.~\ref{fig_RingNetExp}, $\rho_{up}$ and $\rho_{down}$, are obtained for $4$ small resistor networks (i.e., $N = 6,\,12,\,18,$ and $24$) in ring-like arrangements with practically identical resistors of $1\,k\Omega$. The resistor's magnitudes are $1\%$ accurate (according to the manufacturer), which is equivalent to an uncertainty of approximately $10\,\Omega$ [Ohm] per resistor composing the network. We use an ohmmeter to measure the resistance distance between all pairs of nodes and compare the resultant measurements with the exact value found from Eq.~\eqref{eq_ResistDist}, which is the same as the one calculated from Kirchhoff's circuit laws considering all serial and parallel paths. Assuming $1\%$ identical uncertainties for all the $1\,k\Omega$ resistors in the theoretical calculations of Eq.~\eqref{eq_ResistDist}, we compute the uncertainty in $\rho(i,j)$ by error propagation, which also results in a $1\%$ uncertainty for all $\rho(i,j)$ values. Figure~\ref{fig_RingNetExp} shows that the ohmmeter values -- represented by filled (red online) circles -- and $\rho$ values calculated from Eq.~\eqref{eq_ResistDist} -- filled black squares -- are coincident for the $4$ networks (curves superimpose). As expected, these measures coincide because the ohmmeter measures the equivalent resistance between any two point in a circuit, which is identical to the resistance distance measure \cite{Rubido2013}. The negligible differences between these two values (for any pair of nodes) come from the uncertainties associated with each resistor composing the network, which can only be assumed approximately constant for the theoretical calculations.

\begin{figure}[htbp]
    \centering
    \includegraphics[width=0.6\columnwidth]{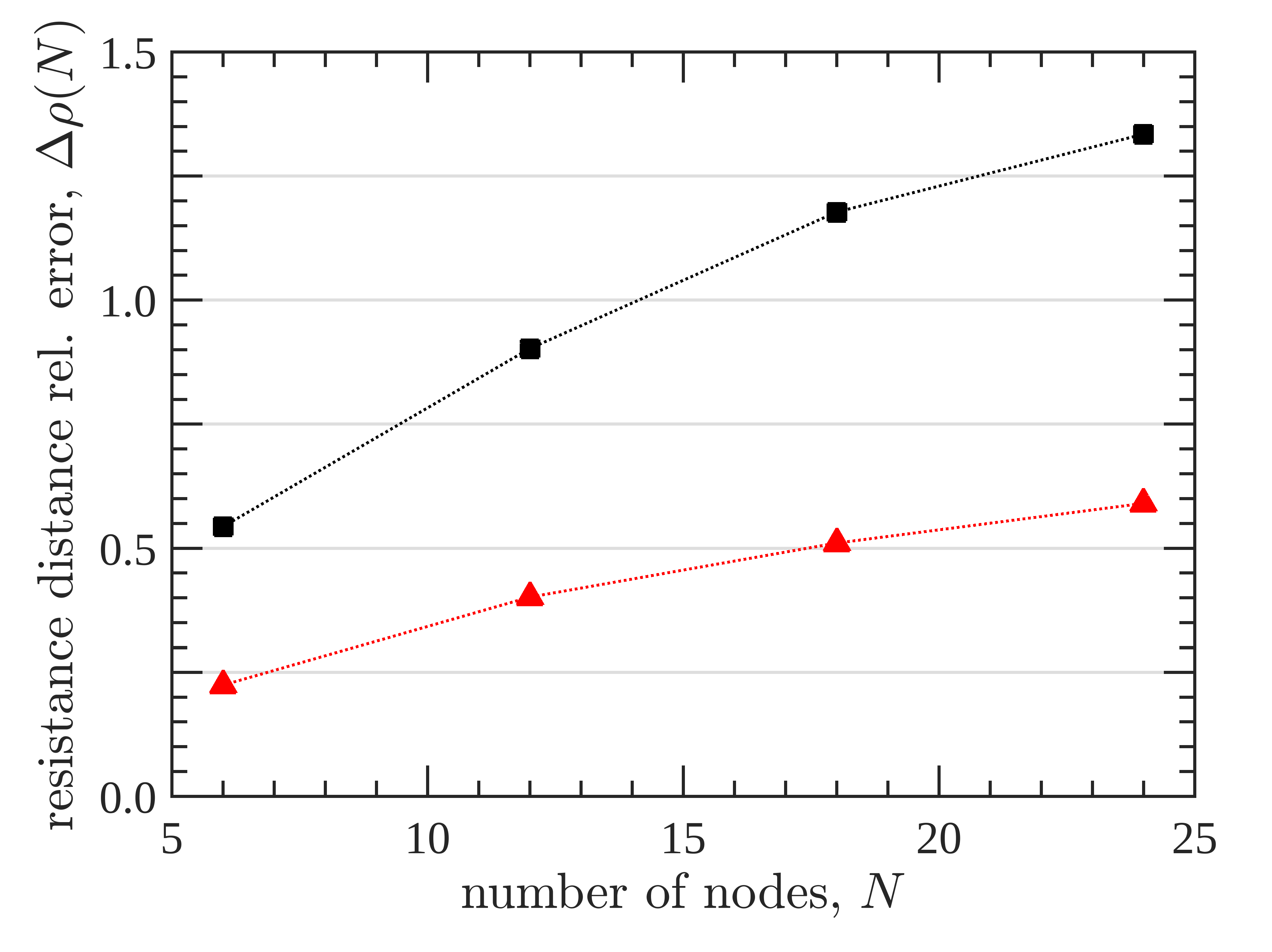}\vspace{-0.5pc}
    \caption{{\bf Average relative error between the approximated resistance distance and the measured experimental values}. The resistance distance average relative error, $\Delta\rho(N) = \frac{2}{N(N - 1)} \sum_{i = 1}^N \sum_{j > i}^N \left|1 - \frac{\rho_{\alpha}(i,j)}{\rho_{e}(i,j)}\right|$, for the upper bound, $\alpha = up$ (filled squares), and for the approximated value, $\alpha = approx$ [Eq.~\eqref{eq_ApproxResistDist}] (filled triangles), is shown with respect to the exact value, $\rho_{e}$. These relative errors correspond to the networks from Fig.~\ref{fig_RingNetExp}.}
    \label{fig_RingNetExpErrors}
\end{figure}

In order to quantify how closely $\rho_{up}$ and $\rho_{approx}$ are to the exact resistance distance value, $\rho_{e}$, from Eq.~\eqref{eq_ResistDist}, we compute the \emph{average relative error}, $\Delta\rho$, as
\begin{equation}
    \Delta\rho(N) = \frac{2}{N(N - 1)} \sum_{i = 1}^N \sum_{j > i}^N \left|1 - \frac{\rho_{\alpha}(i,j)}{\rho_{e}(i,j)}\right|,
    \label{eq_DistRelError}
\end{equation}
where $\alpha = up$ or $approx$. We note from Fig.~\ref{fig_RingNetExpErrors} that, $\Delta\rho$ for the resistor networks of Fig.~\ref{fig_RingNetExp} appears to tend to an asymptotic value as $N$ is increased. This asymptotic value for the upper bound [approximation], $\rho_{up}$ [$\rho_{approx}$], seems to reach a relative error of $100\%$ [$50\%$] with respect to the exact value. The reason behind this large deviation is that, $\rho_{up}$ can surpass almost twofold the exact $\rho_{e}$ value (see the first sets of edges in Fig.~\ref{fig_RingNetExp}). On the other hand, the approximation manages to reduce this error by half. Overall, these resistor networks show that in practical applications we can use the eigenvalue spectra to obtain an approximate estimate of the resistance distance values. As we show in what follows, the relative error decreases significantly for more complex networks, implying that the poor performance of the approximate resistance distance in these resistor networks are due to their particularly regular topology.

\begin{figure}[htbp]
    \centering
    \includegraphics[width=0.49\columnwidth]{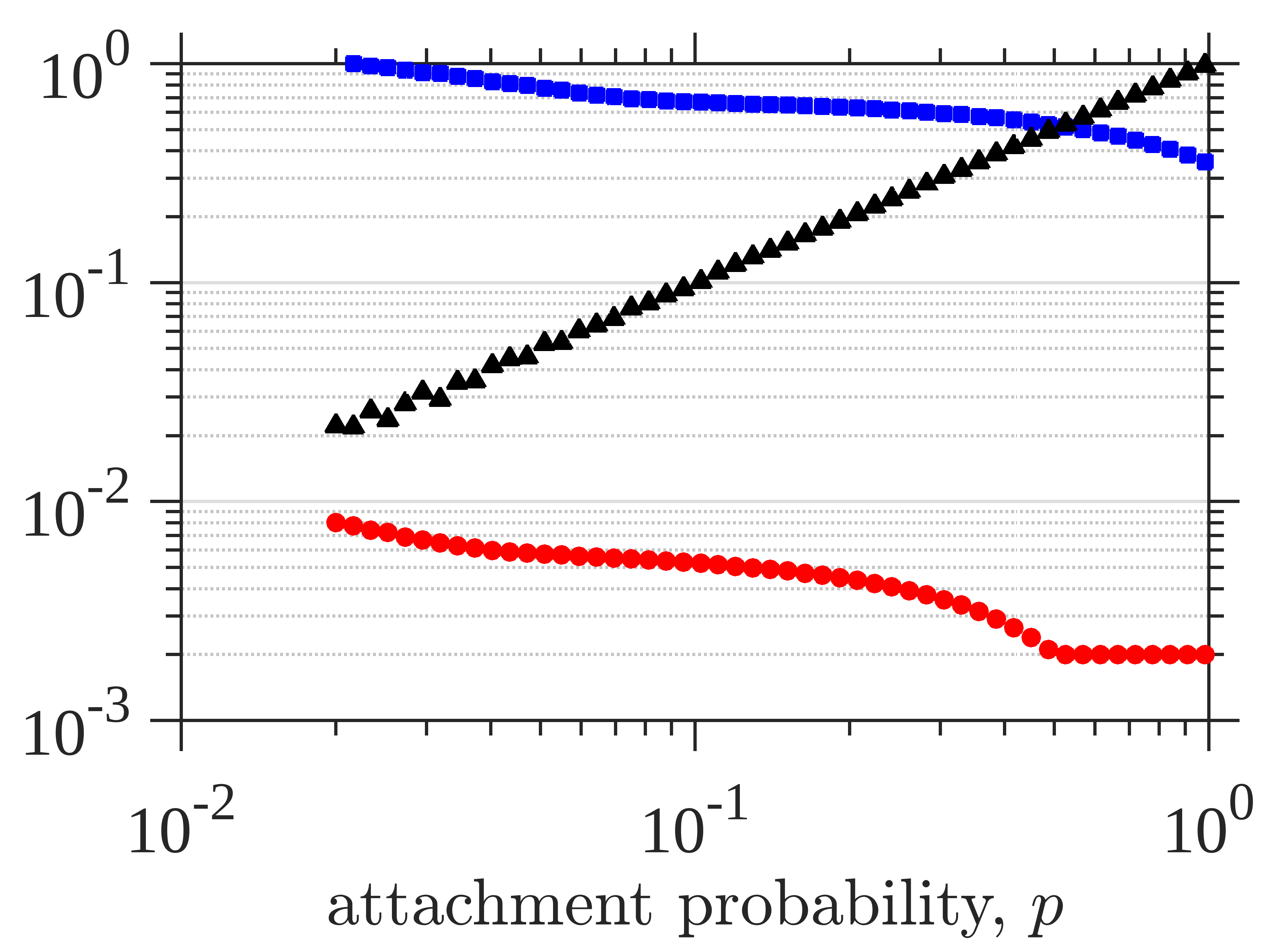}
    \includegraphics[width=0.49\columnwidth]{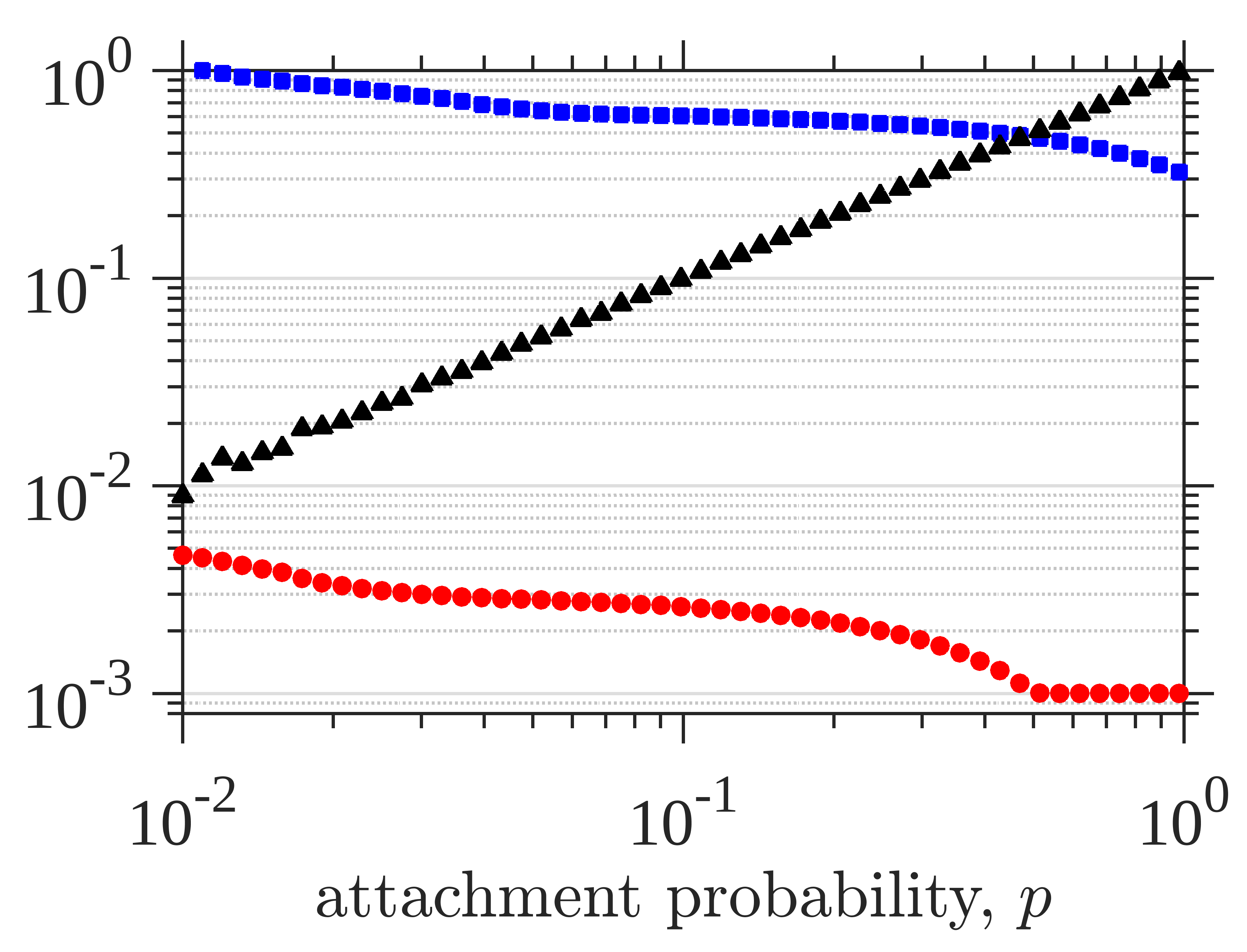}
    \caption{{\bf Erd{\"o}s-R{\'e}nyi network properties as a function of the attachment probability, $p$}. Panels show the normalised average shortest-path, $\frac{\left\langle L \right\rangle(p)}{\left\langle L \right\rangle(p_c)}$, average clustering coefficient, $\frac{\left\langle C \right\rangle(p)}{\left\langle C \right\rangle(1)}$, and resistance distance relative error, $\Delta\rho$, in blue (online) squares, black triangles, and red (online) circles, respectively. These random networks \cite{Erdos1960} are defined by the attachment probability, $p$ (starting at $p \gtrsim p_c = \ln(N)/N$), and the number of nodes, $N$, where $N = 5\times10^2$ and $N = 10^3$ for the left and right panels, respectively. These panels are constructed from the average over $10$ realisations per $p$.}
    \label{fig_RandomNets}
\end{figure}

\begin{figure}[htbp]
    \centering
    \includegraphics[width=0.49\columnwidth]{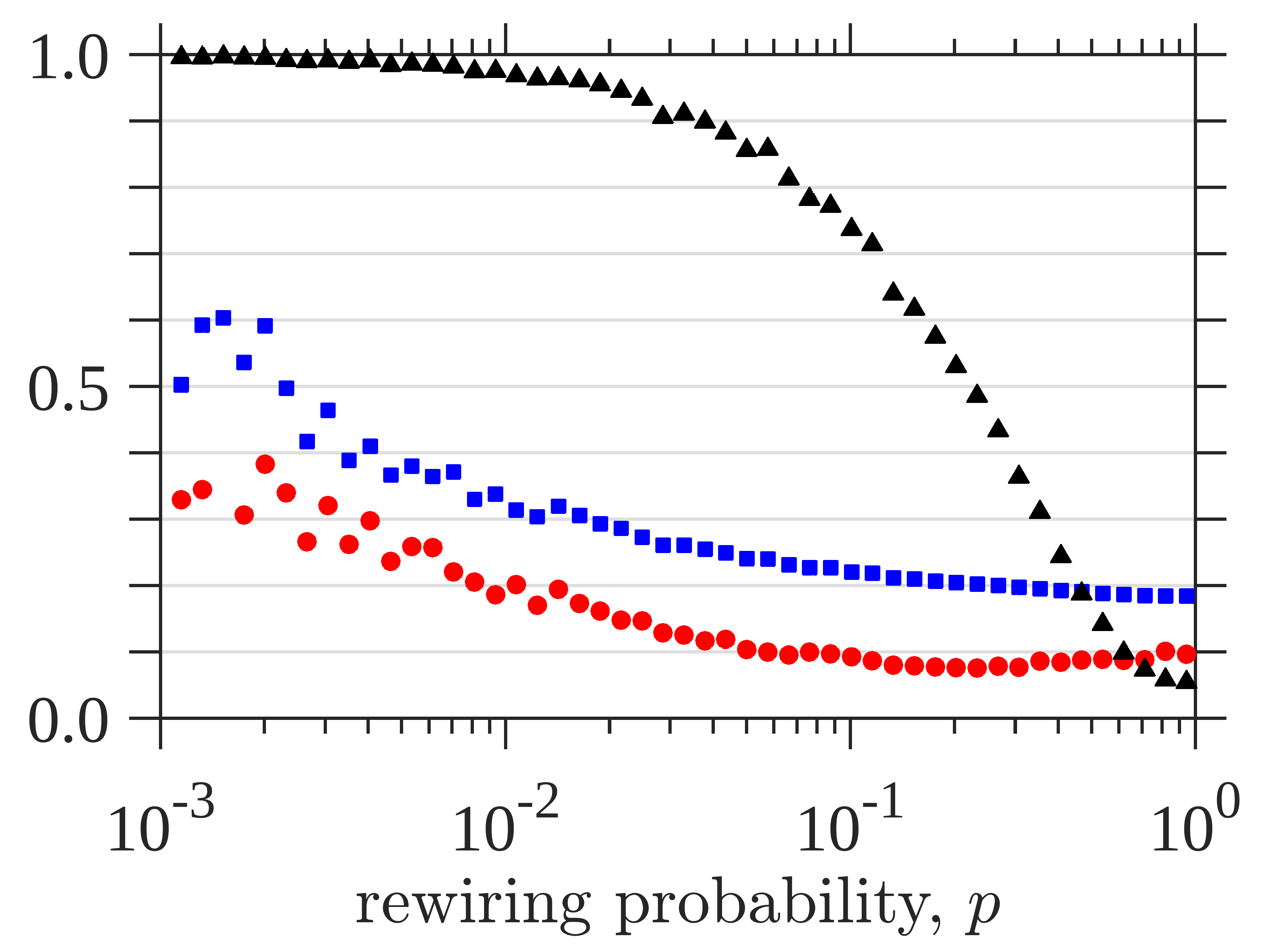}
    \includegraphics[width=0.49\columnwidth]{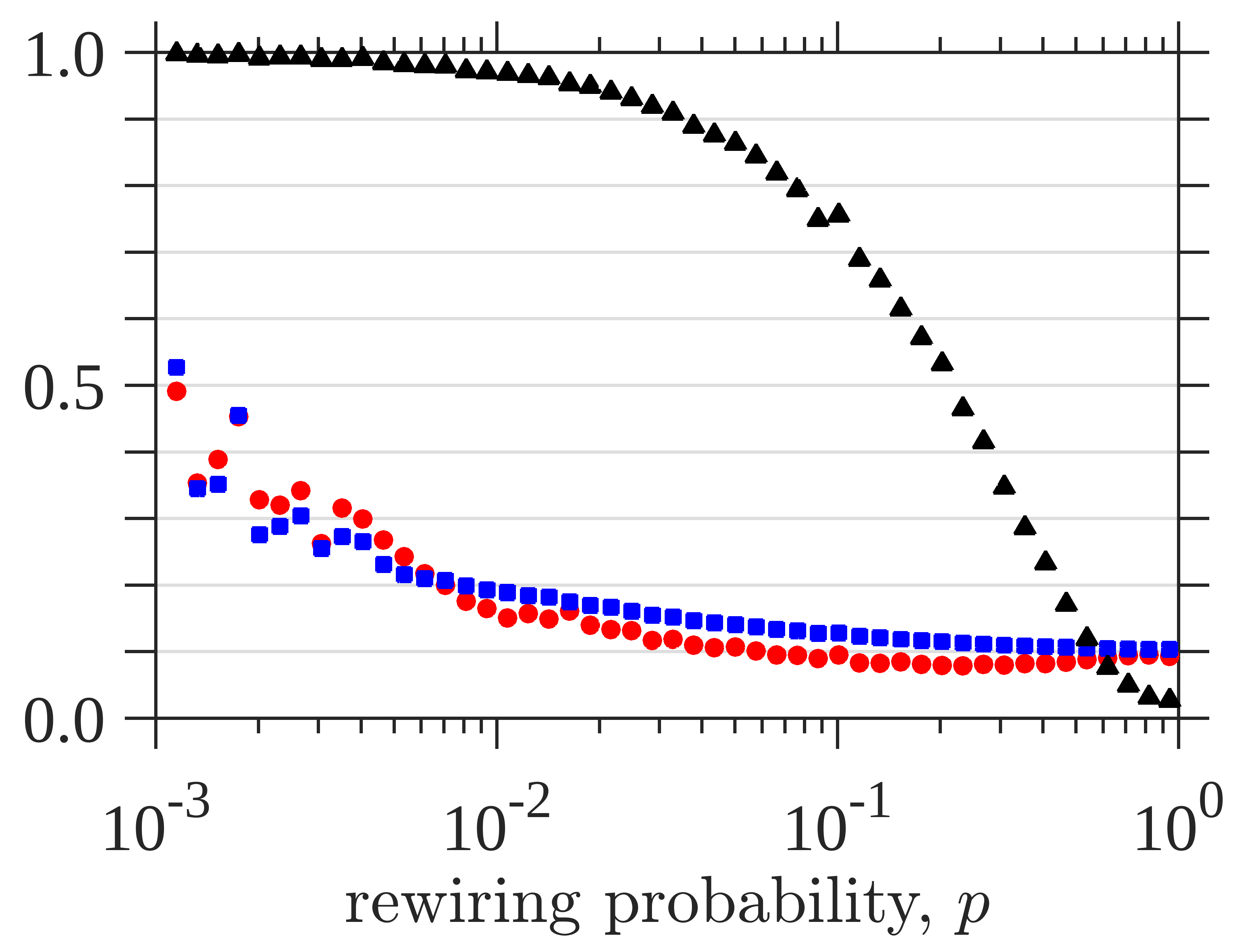}
    \caption{{\bf Watts-Strogatz network properties as a function of the rewiring probability, $p$}. Panels show the normalised average shortest-path, $\frac{\left\langle L \right\rangle(p)}{\left\langle L \right\rangle(0)}$, average clustering coefficient, $\frac{\left\langle C \right\rangle(p)}{\left\langle C \right\rangle(0)}$, and resistance distance relative error, $\Delta\rho$, in blue (online) squares, black triangles, and red (online) circles, respectively. These networks \cite{Strogatz1998} are defined by an initial regular node degree, $K$, which is set to $K = 10$, rewiring probability, $p$, and the number of nodes, $N$, where $N = 5\times10^2$ and $N = 10^3$ in the left and right panels, respectively. Small-world characteristics emerge for $0 < p \lesssim 10^{-1}$.}
    \label{fig_SmallWorldNets}
\end{figure}

In order to quantify the effectiveness of the resistance distance approximation, $\rho_{approx}$, in larger networks, we do the same analysis to random networks \cite{Erdos1960} and small-world networks \cite{Strogatz1998}. For both networks we compute the normalised average shortest-path, $\left\langle L \right\rangle(p)/\max\{\left\langle L \right\rangle(p)\}$, average clustering coefficient, $\left\langle C \right\rangle(p)/\max\{\left\langle C \right\rangle(p)\}$, and resistance distance relative error, $\Delta\rho(p)$, as a function of the attachment or rewiring probability, $p$ (for random networks or small-world networks, respectively). The resultant measures are shown in Figs.~\ref{fig_RandomNets} and \ref{fig_SmallWorldNets} by the filled blue (online) squares, black triangles, and red (online) circles, respectively. The normalised average shortest-paths and average clustering coefficient are shown to identify the rewiring probability region where small-world properties emerge \cite{Strogatz1998}; namely, large average clustering with small average shortest-paths. This region can be seen in Fig.~\ref{fig_SmallWorldNets}, where the small-world characteristics emerge for $10^{-3} \lesssim p \lesssim 10^{-1}$.

We note from Fig.~\ref{fig_RandomNets} that random networks of $N = 500$ (left panel) and $1000$ (right panel) nodes have particularly small resistance distance relative errors, $\Delta\rho \leq 1\%\,\forall\,p$, for all attachment probabilities, $p$. In particular, for each $p$, we compute $\Delta\rho$ from Eq.~\eqref{eq_DistRelError}, where we compare the exact resistance distance value, $\rho_{e}(i,j)$, for each pair of nodes, $(i,j)$, with the approximation given by Eq.~\eqref{eq_ApproxResistDist}, $\rho_{approx}(i,j)$, averaging over all pairs. We observe that as the random network gets more connected, starting from a giant component at $p_c = \ln(N)/N$ and up to the complete graph at $p = 1$, the average shortest-path (blue squares) decreases steadily, non-trivially (it forms a broken monotonically decreasing curve), but only slightly ($1$ order of magnitude less than $\left\langle L(p_c) \right\rangle$), and that the average clustering coefficient (black triangles) increases as a power-law (increasing $2$ orders of magnitude from $\left\langle C(p_c) \right\rangle$). More importantly, as the random network gets more connected, the distance between $\rho_{approx}$ and $\rho_{e}$ [Eq.~\eqref{eq_ResistDist}] decreases (in average) steadily below $1\%$. Overall, we can conclude that random networks have a resistance distance measure that is excellently approximated by Eq.~\eqref{eq_ApproxResistDist}.

On the other hand, from Fig.~\ref{fig_SmallWorldNets} we observe that small-world networks of $N = 500$ (left panel) and $1000$ (right panel) nodes -- with initial regular node degree $k = 10$ \cite{Strogatz1998}-- have an overall higher $\Delta\rho$ than the random networks. We particularly note this difference in the small-world region, where $\Delta\rho$ only decreases below $10\%$ after the rewiring probability, $p$, is higher than the one needed for having small-world characteristics. This ending behaviour can be expected, since for $p = 1$ the Watts-Strogatz model is also a completely random network. The main difference between random networks and small-world networks is the degree regularity. This stems from the initial network, having a uniform node degree of $K = 10$ at $p = 0$, which is maintained for small rewiring probabilities, $p \lesssim 10^{-2}$. In particular, we can see that $\Delta\rho(p)$ tends to $50\%$ for $p\simeq10^{-3}$, as in the resistor networks of Fig.~\ref{fig_RingNetExp} which are also nearly regular graphs.

Now we derive from Eq.~\eqref{eq_Tao} the network's \emph{eigenvector centrality}, $\vec{\psi}^{(c)} = \vec{\psi}_N = \{\psi_1^{(c)},\ldots,\psi_N^{(c)}\}$, which is the Perron-Frobenius vector associated to the maximum eigenvalue \cite{Chang2008} of the adjacency matrix, $\max_{n = 1,\ldots,N}\{\lambda_n(\mathbf{A})\} = \lambda^{(c)}$. This means that $\vec{\psi}^{(c)}$'s components are non-negative, i.e., $\mathbf{A}\vec{\psi}^{(c)} = \lambda^{(c)}\vec{\psi}^{(c)}$ with $[\vec{\psi}^{(c)}]_i = \psi^{(c)}_i\geq0\;\forall\,i$. Thus, we can now use Eq.~\eqref{eq_Tao} to find the eigenvector centrality components directly from the eigenvalue spectra of the adjacency matrix as
\begin{equation}
    \left|[\vec{\psi}^{(c)}]_i\right|^2 = \left|[\vec{\psi}_N]_i\right|^2 =\frac{ \prod_{k = 1}^{N-1} \left[ \lambda_N(\mathbf{A}) - \lambda_k(\mathbf{M}_i)\right] }{ \prod_{k = 1}^{N-1} \left[ \lambda_N(\mathbf{A}) - \lambda_k(\mathbf{A})\right] },
    \label{eq_EigVecCen}
\end{equation}
where we assume (without loss of generality) that the eigenvalue spectra ordering is non-decreasing; that is, $\lambda_1(\mathbf{A}) \leq \lambda_2(\mathbf{A}) \leq \cdots \leq \lambda_N(\mathbf{A})$. Consequently, Eq.~\eqref{eq_EigVecCen} gives the exact eigenvector centrality measure and it is only using adjacency matrix eigenvalue spectra.

\begin{figure}[htbp]
    \centering
    \includegraphics[width=0.49\columnwidth]{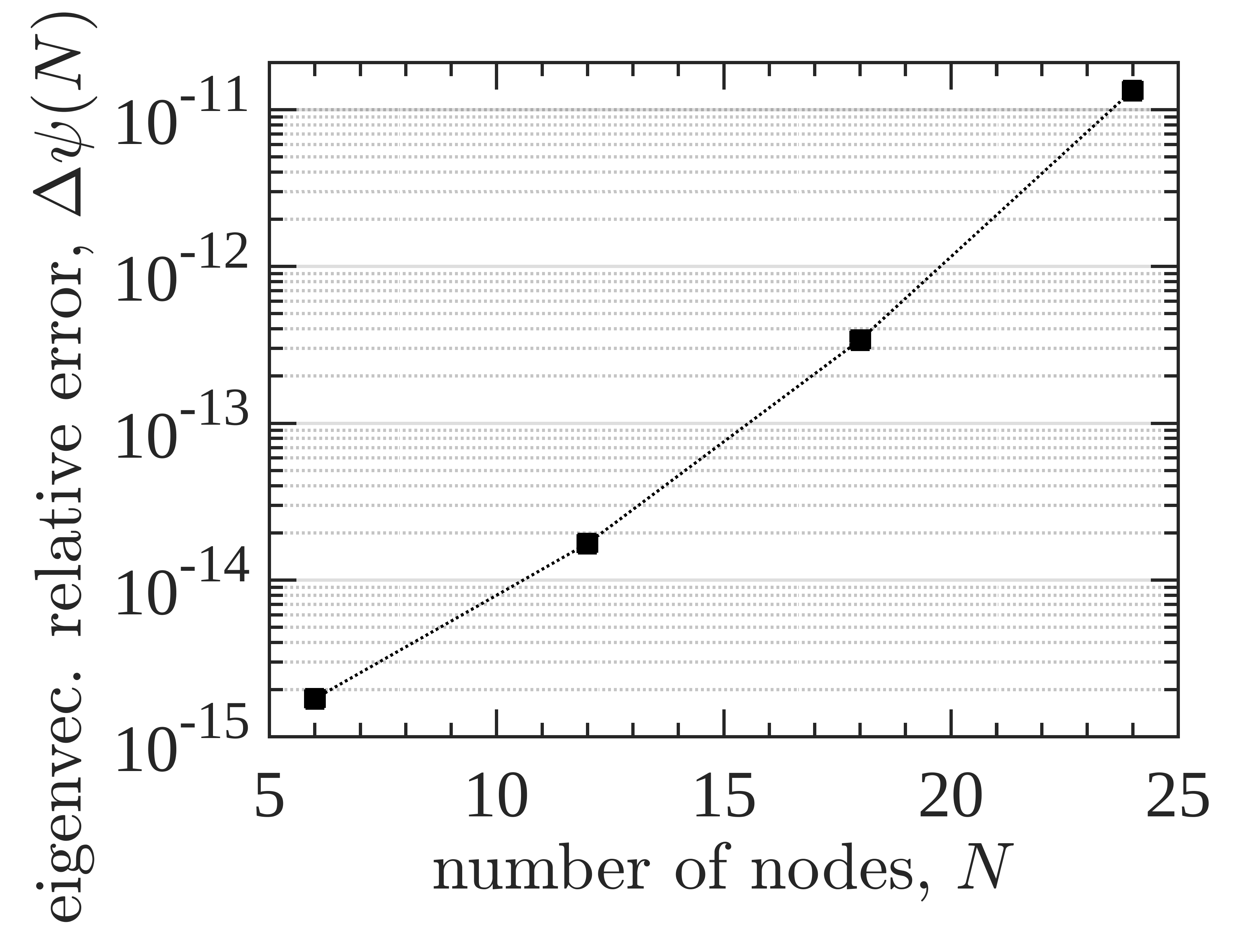}
    \includegraphics[width=0.49\columnwidth]{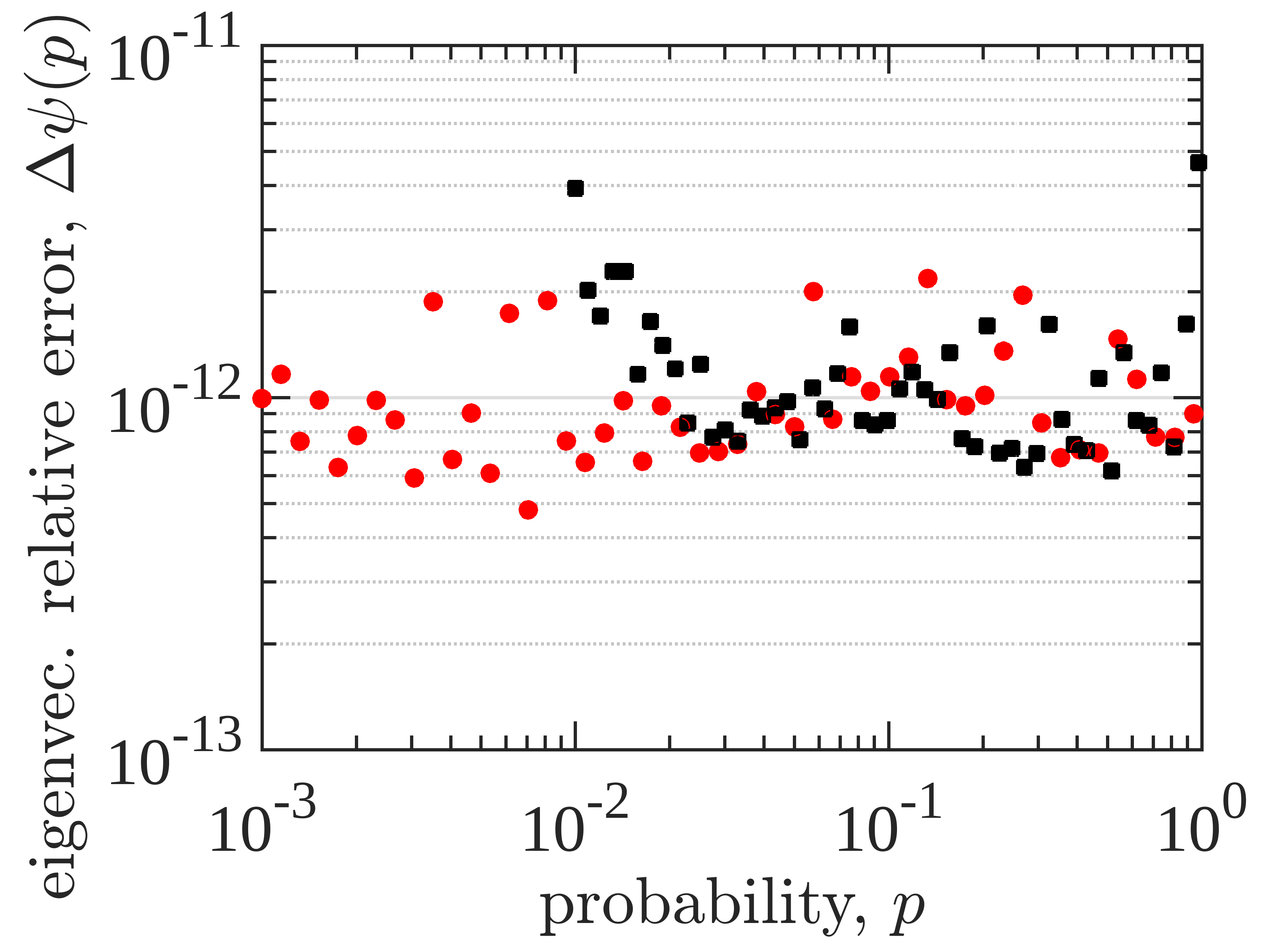}
    \caption{{\bf Average relative error for the eigenvector centrality measure for different networks}. The average relative error, $\Delta\psi = \frac{1}{N} \sum_{i = 1}^N \left|1 - [\vec{\psi}^{(c)}]_i/[\vec{\psi}_{e}^{(c)}]_i\right|$, is the distance between the numerically computed Perron-Frobenius eigenvector, $\vec{\psi}_{e}^{(c)}$ (exact eigenvector centrality measure), and our analytical derivation of Eq.~\eqref{eq_EigVecCen}, $\vec{\psi}^{(c)}$. In the left [right] panel we show by filled squares [filled squares and circles] $\Delta\psi(N)$ [$\Delta\psi(p)$] for the resistor networks [random networks and small-world networks] in Fig.~\ref{fig_RingNetExp} [Figs.~\ref{fig_RandomNets} and \ref{fig_SmallWorldNets}, respectively].}
    \label{fig_NetEigErrors}
\end{figure}

We test the accuracy of Eq.~\eqref{eq_EigVecCen} by comparing its value for the previous networks with that of the Perron-Frobenius eigenvector directly calculated from the adjacency matrix. Namely, we find the average relative error $\Delta\psi$ and show the resultant values in Fig.~\ref{fig_NetEigErrors}. Specifically, $\Delta\psi$ is the difference between the numerically calculated eigenvector centrality, $\vec{\psi}_{e}^{(c)}$, and the one found from Eq.~\eqref{eq_EigVecCen}, $\vec{\psi}^{(c)}$. On the left panel of Fig.~\ref{fig_NetEigErrors}, we can see that $\Delta\psi(N)$ for the experimental resistor-networks is always within numerical errors (i.e., due to round-off and truncation errors). Similarly, on the right panel of Fig.~\ref{fig_NetEigErrors}, we can see this negligible error also happens when testing Eq.~\eqref{eq_EigVecCen} with the $N = 1000$ nodes random networks of Fig.~\ref{fig_RandomNets} (black squares) and small-world networks of Fig.~\ref{fig_SmallWorldNets} (red circles), which is happening for all attachment and rewiring probabilities, respectively.

\section{Conclusions}\label{sec_Concls}
In this work, we use the eigenvector-eigenvalue identity to express network measures that are based on eigenvectors and/or eigenvalues, only in terms of the eigenvalues. Although our works focuses on the resistor distance and eigenvector centrality measures, it is unrestricted to these measures, being applicable to any measure that requires knowing eigenvector sets.

On the one hand, we derive an expression to approximate the resistance distance, $\rho_{approx}$ [Eq.~\eqref{eq_ApproxResistDist}], values of a network from its Laplacian matrix eigenvalue spectra. We first use experimentally implemented, small-sized, resistor networks to explain how and why our approximation works [see Fig.~\ref{fig_RingNetExp}]. We then show how efficiently our approximation matches the exact resistance distance values by testing it on synthetically generated random and small-world networks [Figs.~\ref{fig_RandomNets} and \ref{fig_SmallWorldNets}]. From these numerical experiments, we conclude that $\rho_{approx}$ in random networks typically misses the exact value by less than $1\%$ (in average), regardless of the attachment probability [see Fig.~\ref{fig_RandomNets}]. For small-world networks, the approximation misses by $50\%$ when the network is almost regular ($p \simeq 0$), decreasing its relative error steadily from $50\%$ to $10\%$ during the small-world region as edges are rewired [see Fig.~\ref{fig_SmallWorldNets}]. We can explain this behaviour due to the inherent regularity of small-world networks, which inherits the initial regular structure that creates degenerate eigenvalues, making the eigenvector-eigenvalue identity to fail. Consequently, we expect that our $\rho_{approx}$ will always work better for networks that have an heterogeneous (and broad) degree distribution -- such as those from real-world complex systems -- than for those with a narrow degree distribution.

On the other hand, we derived an exact expression for the eigenvector centrality of any network from the eigenvalue spectra of its adjacency matrix. We show that the differences between our derived eigenvector expression and the numerically found eigenvector centrality measure are negligible in all cases studied -- resistor, random, and small-world networks [see Fig.~\ref{fig_NetEigErrors}]. This means that in general, when eigenvalues are non-degenerate, our expression allows to find the eigenvector centrality measure without the need to find the adjacency matrix eigenvectors, which is computationally demanding for large-sized networks. Consequently, and given the current relevance that the eigenvector centrality measure is having in current biomedical research (e.g., in Network Neuroscience), we believe our expression will become increasingly useful.

\begin{acknowledgments}
C.G. acknowledges funds from the Agencia Nacional de Investigaci{\'o}n e Innovaci{\'o}n (ANII), Uruguay, POS\_NAC\_2018\_1\_151237. J.G. acknowledges funds from the ANII, Uruguay, POS\_NAC\_2018\_1\_151185. All authors acknowledge the Comisi{\'o}n Sectorial de Investigaci{\'o}n Cient{\'i}fica (CSIC), Uruguay, group grant ``CSIC2018 - FID13 - grupo ID 722''.
\end{acknowledgments}


\end{document}